# On the Achievable Improvement by the Linear Minimum Mean Square Error Detector

Manar Mohaisen, *Student Member, IEEE*, and KyungHi Chang, *Senior Member, IEEE*
The Graduate School of IT and T, Inha University
253 Yonghyun-Dong, Nam-Gu, 402-751 Incheon, KOREA
Email: lemanar@hotmail.com, khchang@inha.ac.kr

*Abstract*—Linear minimum mean square error (MMSE) detector has been shown to alleviate the noise amplification problem, resulting in the conventional zero-forcing (ZF) detector. In this paper, we analyze the performance improvement by the MMSE detector in terms of the condition number of its filtering matrix, and in terms of the post-precessing signal to noise ratio (SNR) improvement. To this end, we derive explicit formulas for the condition numbers of the filtering matrices and the post-processing SNRs. Analytical and simulation results demonstrate that the improvement achieved by the MMSE detector over the ZF detector is not only dependent on the noise variance and the condition number of the channel matrix, but also on how close the smallest singular values are to the noise variance.

## I. INTRODUCTION

Multiple-input multiple-output spatial multiplexing (MIMO-SM) techniques are considered as the key technology to increase the channel capacity in mobile communication systems beyond third-generation (3G) [1]. In MIMO-SM, signals are transmitted simultaneously from different spatial elements, i.e., antennas, using the same time-frequency resources [2]. Therefore, the channel capacity is increased without requiring any additional spectral resources, which are not only expensive but also scarce. Indeed, the attained capacity by the MIMO multiplexing techniques depends crucially on the detection algorithm which is employed at the receiver side to demultiplex the transmitted signals.

Maximum-likelihood (ML), which employs brute-force search, is the optimum detection technique for the MIMO multiplexing systems [3]. Nonetheless, its exponential complexity is inapplicable in computational complexity and latency limited mobile communication systems.

Several suboptimal detection schemes were proposed in the literature [4]-[7]. It is clearly shown in several works that taking the noise into consideration increases the detection efficiency (see for instance [8] and references therein). This improvement in the efficiency can be translated into reduction in the computational complexity, error enhancement, or both of them. Nonetheless, this improvement achieved by the minimum-mean square error (MMSE) criterion is referred to the value of the noise variance, or more accurately to the signal to noise ratio (SNR). That is; when the noise variance is small, the performances of the ZF and MMSE linear detectors coincide [9]. We show that this reasoning, which is frequently used in the literature, is inaccurate.

**Contributions.** In this paper, we investigate the improvement achieved by the linear detector using the MMSE criterion over that using the ZF criterion. The performance improvement is given as the ratio between the *condition numbers* of the filtering matrices. We show that the achieved improvement by the MMSE criterion does not only depend on the noise variance, but also on how close the noise variance is to the singular values of the channel matrix. Also, we derive the post-processing SNRs of the linear ZF and MMSE detectors and show the superiority of the detection algorithm when noise is considered in constructing the filtering matrix.

The rest of this paper is as follows. In Section II, we introduce the system model and give a brief review of linear detection schemes. In Section III, we analyze the condition numbers of the filtering matrices of the ZF and MMSE detectors. Analytical and simulation results are shown in Section IV, and conclusions are drawn in Section V.

We define now some variables and notations used throughout the paper. The notation $\text{Tr}(\cdot)$ denotes the trace of a matrix, $\text{E}(\cdot)$ denotes the expectation, and $(\cdot)^H$ is the Hermitian transpose. $\lambda_i(\mathbf{H})$ and $\sigma_i(\mathbf{H})$ are the $i$-th eigenvalue and singular value of the matrix $\mathbf{H}$, respectively.

## II. SYSTEM MODEL AND REVIEW OF THE LINEAR DETECTION TECHNIQUES

We consider a MIMO spatial multiplexing system with $n_T$ transmit and $n_R$ receive antennas. For the sake of simplicity, we assume that $N = n_T = n_R$. The received vector $\mathbf{r} \in \mathbb{C}^N$ is given by:

$$\mathbf{r} = \mathbf{H}\mathbf{x} + \mathbf{n}, \tag{1}$$

where $\mathbf{n}$ is the Gaussian noise vector with $\text{E}[\mathbf{n}\mathbf{n}^H] = \sigma_n^2 \mathbf{I}$, and $\mathbf{H}$ is the channel matrix whose element $h_{i,j}$ is the zero-mean unit-variance complex Gaussian random variable. Also, $\mathbf{x}$ is the transmitted data vector whose elements are drawn independently from a quadratic amplitude modulation (QAM) set.

Linear detection algorithms are the simplest approaches used to solve (1). The main idea behind the linear detectors is to treat the received vector by a filtering matrix $\mathbf{W}$, which is constructed using a performance-based criterion. The zero-forcing and minimum-mean square error criteria are used in

This work was supported by the Korea Science and Engineering Foundation (KOSEF) grant funded by the Korea government (MOST) (No. R01-2008-000-20333-0).

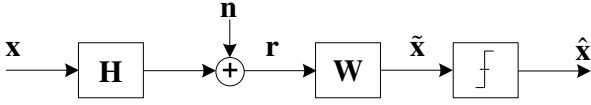

Fig. 1. MIMO spatial multiplexing with linear receiver.

the linear ZF and MMSE detection algorithms, respectively. Hence, the filtering matrices for the ZF and MMSE detectors are given as follows:

$$\mathbf{W}_{\text{zf}} = \left(\mathbf{H}^H\mathbf{H}\right)^{-1}\mathbf{H}^H \quad (2)$$

and

$$\mathbf{W}_{\text{mmse}} = \left(\mathbf{H}^H\mathbf{H} + \sigma_n^2\mathbf{I}_N\right)^{-1}\mathbf{H}^H, \quad (3)$$

where $\sigma_n^2$ is the noise variance and $\mathbf{I}_N$ is the $N \times N$ identity matrix. Fig. 1 depicts the MIMO multiplexing system with linear receiver. In the sequel, the channel matrix $\mathbf{H}$ is considered to be perfectly known at the receiver.

## III. ACHIEVED IMPROVEMENT BY THE MMSE DETECTOR

### A. Improvement in the Conditionality of the Filtering Matrix

In this section, we evaluate the improvement achieved by the linear MMSE over the linear ZF algorithm in terms of the condition number of their corresponding filtering matrices. Note that the condition number is used to demonstrate the accuracy of the recovered signal, and to show the system's immunity against channel estimation errors. The condition number of an $M \times N$ matrix $\mathbf{A}$ is defined as the ratio between its maximal and minimal singular values; that is,

$$\text{cond}(\mathbf{A}) = \|\mathbf{A}\| \cdot \|\mathbf{A}^{-1}\| = \frac{\sigma_{\max}(\mathbf{A})}{\sigma_{\min}(\mathbf{A})}, \quad (4)$$

where $\|\mathbf{A}\|$ is the Euclidean norm of the matrix $\mathbf{A}$, which equals the maximum singular value of $\mathbf{A}$. To proceed in the analysis, we introduce the following Inequality and Lemma.

**Inequality III.1.** *Weyl's Inequalities (Eigenvalues)* [10]: *Let $\Sigma$ and $\Delta$ be $N \times N$ Hermitian matrices, $\lambda_1(\Sigma) \geq \lambda_2(\Sigma) \geq \cdots \geq \lambda_N(\Sigma) > 0$, $\lambda_1(\Delta) \geq \lambda_2(\Delta) \geq \cdots \geq \lambda_N(\Delta) > 0$, and $\lambda_1(\Gamma) \geq \lambda_2(\Gamma) \geq \cdots \geq \lambda_N(\Gamma) > 0$ are the eigenvalues for $\Sigma$, $\Delta$, and $\Gamma = (\Sigma + \Delta)$, respectively. Then,*

$$\lambda_i(\Gamma) \geq \begin{cases} \lambda_i(\Sigma) & + \lambda_N(\Delta) \\ \lambda_{i+1}(\Sigma) & + \lambda_{N-1}(\Delta) \\ & \vdots \\ \lambda_N(\Sigma) & + \lambda_i(\Delta). \end{cases} \quad (5)$$

**Lemma III.2.** *Let $\mathbf{A}$ be an $N \times N$ Gram matrix, i.e., $\mathbf{A} = \mathbf{B}^H\mathbf{B}$ for any nonsingular matrix $\mathbf{B}$. Then, $\lambda_i(\mathbf{A}) > 0$ and $\lambda_i(\mathbf{A})$ is also a singular value.*

*Proof:* By definition, $\sigma_i(\mathbf{H}) = \lambda_i^{1/2}(\mathbf{H}^H\mathbf{H})$, which implies that $\sigma_i^2(\mathbf{H}) = \lambda_i(\mathbf{H}^H\mathbf{H})$. Notice that the singular value decomposition (SVD) of $(\mathbf{H}^H\mathbf{H}) = \mathbf{U}\Sigma^H\mathbf{V}^H\mathbf{V}\Sigma\mathbf{U}^H = \mathbf{U}(\Sigma^H\Sigma)\mathbf{U}^H$, which is equivalent to the eigenvalue decomposition with orthonormal eigenvectors. Therefore, $\sigma_i^2(\mathbf{H}) = \sigma_i(\mathbf{H}^H\mathbf{H}) = \lambda_i(\mathbf{H}^H\mathbf{H})$. ∎

From Weyl's inequalities for the eigenvalues and Lemma (III.2), it follows that:

$$\sigma_i(\Gamma) \geq \begin{cases} \sigma_i(\Sigma) & + \sigma_N(\Delta) \\ \sigma_{i+1}(\Sigma) & + \sigma_{N-1}(\Delta) \\ & \vdots \\ \sigma_N(\Sigma) & + \sigma_i(\Delta). \end{cases} \quad (6)$$

Therefore,

$$\sigma_1(\Gamma) \geq \sigma_1(\Sigma) + \sigma_N(\Delta) \quad (7)$$

and

$$\sigma_N(\Gamma) \geq \sigma_N(\Sigma) + \sigma_N(\Delta). \quad (8)$$

Now, we define the Gram matrices $\Sigma = \mathbf{H}^H\mathbf{H}$ and $\Delta = \Lambda^H\Lambda$, where $\Lambda = \sigma_n\mathbf{I}_N$. Then, dividing both sides of (7) by the corresponding sides of (8) results in:

$$\frac{\sigma_1(\Gamma)}{\sigma_N(\Gamma)} = \text{cond}(\Gamma),$$
$$\approx \frac{\sigma_1(\Sigma) + \sigma_N(\Delta)}{\sigma_N(\Sigma) + \sigma_N(\Delta)},$$
$$= \frac{\text{cond}(\Sigma) + \sigma_n^2/\sigma_N(\Sigma)}{1 + \sigma_n^2/\sigma_N(\Sigma)}. \quad (9)$$

**Lemma III.3.** *For any nonsingular $N \times N$ matrix $\mathbf{A}$, $\text{cond}(\mathbf{A}) = \text{cond}(\mathbf{A}^{-1})$.*

*Proof:* The proof of this lemma follows from the fact that $\sigma_1(\mathbf{H}^{-1}) = 1/\sigma_N(\mathbf{H})$ and $\sigma_N(\mathbf{H}^{-1}) = 1/\sigma_1(\mathbf{H})$. ∎

Then, the ratio between the condition number of the filtering matrices of the linear MMSE and ZF detectors is approximated as follows:

$$\frac{\text{cond}(\mathbf{W}_{\text{mmse}})}{\text{cond}(\mathbf{W}_{\text{zf}})} = \frac{\text{cond}\left(\left(\mathbf{H}^H\mathbf{H} + \sigma_n^2\mathbf{I}_N\right)^{-1}\mathbf{H}^H\right)}{\text{cond}\left(\left(\mathbf{H}^H\mathbf{H}\right)^{-1}\mathbf{H}^H\right)},$$
$$\approx \frac{\text{cond}(\Gamma)}{\text{cond}(\Sigma)},$$
$$= \frac{\text{cond}(\Sigma) + \sigma_n^2/\sigma_N(\Sigma)}{1 + \sigma_n^2/\sigma_N(\Sigma)} \times \frac{1}{\text{cond}(\Sigma)},$$
$$= \frac{1 + \sigma_n^2/\sigma_1^2(\mathbf{H})}{1 + \sigma_n^2/\sigma_N^2(\mathbf{H})}. \quad (10)$$

Based on (10), we make the following remarks:

1) When the channel matrix $\mathbf{H}$ is orthogonal, i.e., $\sigma_1(\mathbf{H}) = \sigma_N(\mathbf{H})$, the condition number of $\mathbf{W}_{\text{zf}}$ equals that of $\mathbf{W}_{\text{mmse}}$.
2) The improvement by the MMSE detector does not depend only on the noise variance $\sigma_n^2$, but also on how close the noise variance is to the minimum singular value $\sigma_N(\mathbf{H})$. Also, it depends on the conditionality of the channel matrix $\mathbf{H}$.

## B. Improvement in the Post-processing SNR

In the case of ZF detector, the post-processing signal vector is given as follows:

$$\hat{\mathbf{x}} = \mathbf{H}^{-1}\mathbf{r} = \mathbf{x} + \mathbf{H}^{-1}\mathbf{n}. \quad (11)$$

The post-processing SNR of the ZF detector is then given by:

$$\begin{aligned}\text{SNR}_{\text{zf}} &= \frac{\text{E}(\mathbf{x}^H\mathbf{x})}{\text{E}(\mathbf{n}^H\mathbf{H}^{-1^H}\mathbf{H}^{-1}\mathbf{n})}, \\ &= \frac{N}{\text{E}(\text{tr}(\mathbf{H}^{-1^H}\mathbf{H}^{-1}\mathbf{n}\mathbf{n}^H))}.\end{aligned} \quad (12)$$

Let $(\mathbf{H}\mathbf{H}^H) = \mathbf{V}\Sigma\mathbf{V}^H$ be the singular value decomposition of the Gramian matrix $(\mathbf{H}\mathbf{H}^H)$, then (12) can be simplified as follows:

$$\text{SNR}_{\text{zf}} = \frac{N}{\sum_{i=1}^{N}\left(\frac{\sigma_n^2}{\sigma_i^2(\mathbf{H})}\right)}, \quad (13)$$

where the expectation in the denominator is only taken with respect to the noise vector $\mathbf{n}$.

Based on the analysis carried out in [11] for the case of multi-user precoding, we conducted the analysis for the linear MMSE detector. The post-processing SNR is given as following:

$$\text{SNR}_{\text{mmse}} = \frac{a+b}{\sigma_n^2(N+1)c + Nb - a}, \quad (14)$$

where

$$a = \left(\sum_{i=1}^{N}\frac{\sigma_i^2(\mathbf{H})}{\sigma_i^2(\mathbf{H}) + \sigma_n^2}\right)^2, \quad (15)$$

$$b = \sum_{i=1}^{N}\left(\frac{\sigma_i^2(\mathbf{H})}{\sigma_i^2(\mathbf{H}) + \sigma_n^2}\right)^2, \quad (16)$$

and

$$c = \sum_{i=1}^{N}\frac{\sigma_i^2(\mathbf{H})}{(\sigma_i^2(\mathbf{H}) + \sigma_n^2)^2}. \quad (17)$$

Notice that for $\sigma_n^2 = 0$, $\text{SNR}_{\text{mmse}} = \text{SNR}_{\text{zf}}$.

## IV. SIMULATION RESULTS AND DISCUSSIONS

In this section, we consider the elements of the channel matrix $\mathbf{H}$ to be independent and identically distributed (i.i.d.) and to follow a zero-mean unit-variance complex Gaussian distribution. Each realization of the channel matrix is normalized such that

$$\sum_{i=1}^{N}\sum_{j=1}^{N}|h_{i,j}|^2 = \sum_{i=1}^{N}\sigma_i^2(\mathbf{H}) = N^2, \quad (18)$$

where $h_{i,j}$ is the row-column element of the channel matrix $\mathbf{H}$. Therefore, the SNR at each receive antenna equals $\frac{N}{\sigma_n^2}$. Also, quadrature phase shift keying (QPSK) modulation is used.

Fig. 2 depicts the gain attained by the MMSE detector over the ZF detector in terms of the post-processing SNR. As $N$ increases, the gain achieved by the MMSE detector increases, attaining about 15dB at a receive SNR of 0dB. Indeed, it is the post-processing SNR of the ZF detector that deteriorates due to the vanishing of the minimum singular value of the channel matrix as $N$ increases.

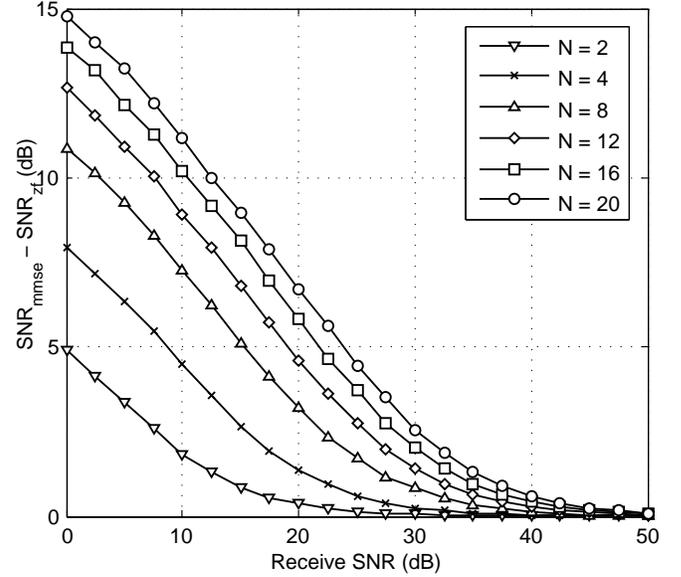

Fig. 2. Achieved gain by the linear MMSE over the ZF detector in terms of the post-processing signal to noise ratio.

TABLE I
MINIMUM SINGULAR VALUE AND CONDITION NUMBER OF THE CHANNEL MATRIX FOR SEVERAL CONFIGURATIONS. VALUES ARE AVERAGED OVER 10,000 INDEPENDENT CHANNEL REALIZATIONS.

| $N$ | 2 | 4 | 8 | 12 | 16 | 20 |
|---|---|---|---|---|---|---|
| $\text{E}(\sigma_N(\mathbf{H}))$ | 0.642 | 0.447 | 0.314 | 0.257 | 0.224 | 0.198 |
| $\text{E}(\text{cond}(\mathbf{H}))$ | 4.27 | 10.82 | 24.32 | 37.82 | 49.85 | 64.95 |

Table I depicts the expectations of the minimum singular value and the condition number of the channel matrix for several $N$. The given values are the average over 10,000 independent realizations of $\mathbf{H}$. As $N$ increases, we remark that minimum singular value decreases, and consequently the condition number increases. This clearly explains the poor performance of the linear ZF detector, where the post-processing noise variance depends directly on the inverse of the singular values. These results coincide with the following theorem (see Thm. 2.36 in [13] and Thm. 5.1 in [14]).

**Theorem IV.1.** *Consider the $N \times N$ standard complex Gaussian matrix $\mathbf{H}$, i.e., $h_{i,j} \sim CN(0,1)$. The minimum singular value of $\mathbf{H}$ satisfies*

$$\lim_{N\to\infty} P[N\sigma_N(\boldsymbol{H}) \geq \text{x}] = \text{e}^{-\text{x}-\text{x}^2/2}. \quad (19)$$

*where $P[x]$ is the probability of $x$.*

Fig. 3 shows the cumulative distribution function (cdf) of the minimum singular value $\sigma_N(\mathbf{H})$ for several $N$ values. As expected from Table I, for an $N \times N$ matrix $\mathbf{A}$ and an $M \times M$ matrix $\mathbf{B}$ with $M > N$, the following is satisfied

$$P[\sigma_M(\mathbf{B}) < x] \geq P[\sigma_N(\mathbf{A}) < x]. \quad (20)$$

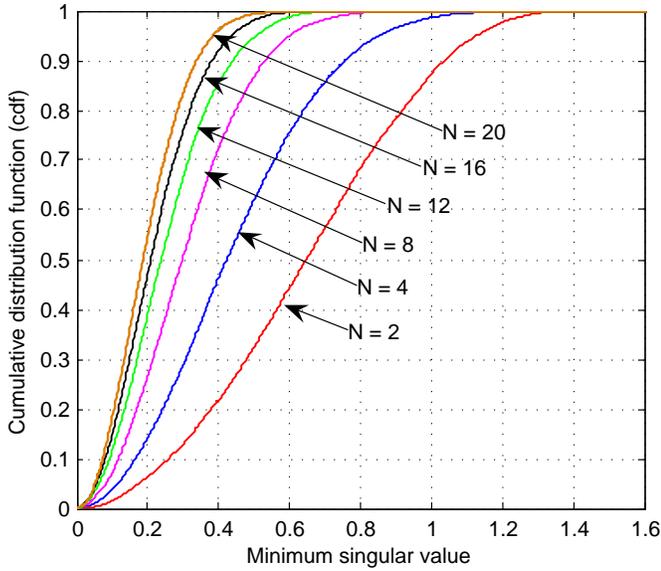

Fig. 3. Cumulative distribution function (cdf) of the minimum singular value of the channel matrix in several system configurations.

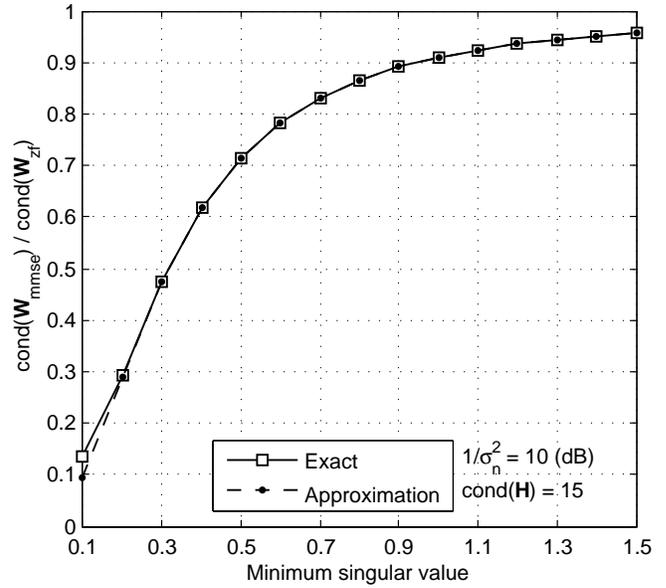

Fig. 5. Effect of the minimum singular value of the channel matrix on the improvement by the MMSE detector in the low SNR regime.

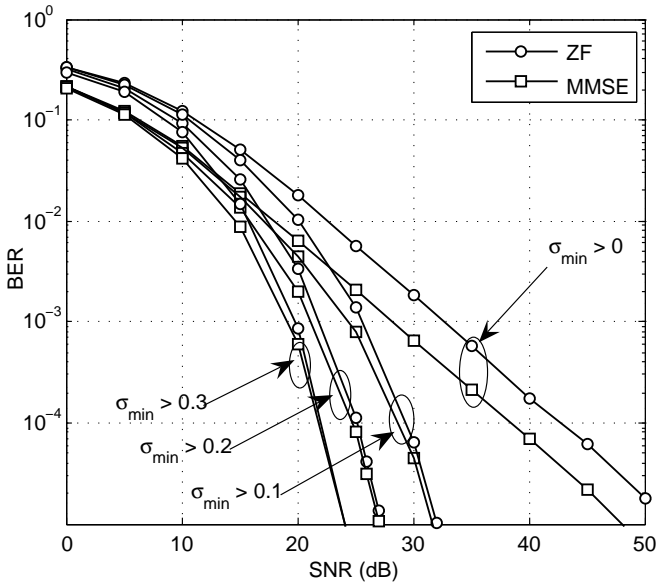

Fig. 4. Effect of the minimum singular value of the channel matrix on the performance of the linear detectors in $4 \times 4$ MIMO multiplexing system.

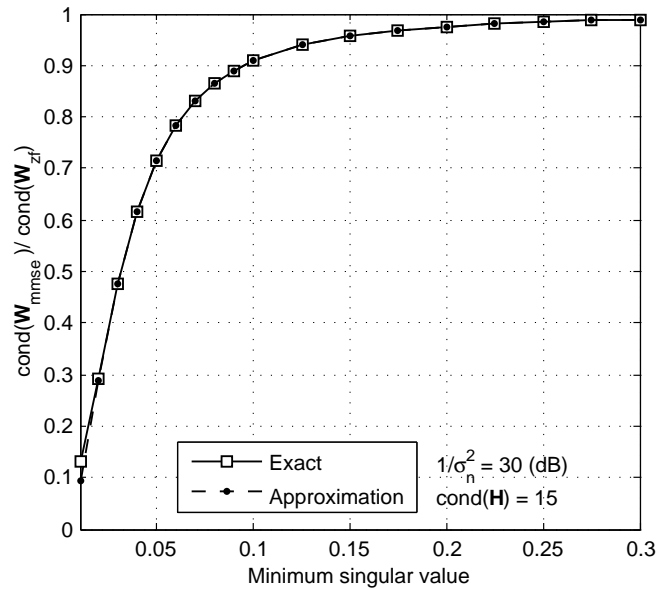

Fig. 6. Effect of the minimum singular value of the channel matrix on the improvement by the MMSE detector in the high SNR regime.

This gives the reason behind the degradation in the performance of the ZF detector as $N$ increases.

Fig. 4 depicts the bit error rate (BER) performance of the linear MMSE and ZF detectors in $4 \times 4$ MIMO multiplexing system. Notice that the power constraint in (18) is still valid for all the obtained simulations. When no constraint is imposed on the minimum singular value, i.e., $\sigma_{\min} > 0$, the MMSE outperforms the ZF detector by about 4.2dB for all the simulated SNR range. Notice that even at SNR of 50dB, viz., $\sigma_n^2 = 2 \times 10^{-5}$, the improvement by the MMSE detector is kept fixed. This indicates that the improvement by the MMSE in the error rate does not only depend on the plain value of the noise variance.

To show the relationship between the minimum singular value and the improvement achieved by the MMSE detector, we assign a lower bound on the minimum singular value of $\mathbf{H}$. We remarked that the BER performances of the ZF and MMSE detectors coincide in the case of ($\sigma_{\min} = 0.3$), ($\sigma_{\min} = 0.2$), and ($\sigma_{\min} = 0.1$) at SNR of 25dB, 28dB, and 33dB, respectively. This is equivalent to $\sigma_{\min}^2/\sigma_n^2$ equals 14, 12.5, and 10, respectively. We don't have a good explanation of these results, and leave it for further studies. This clearly shows that the improvement by the linear MMSE detector depends on the ratio between the minimum singular value of $\mathbf{H}$ and the noise variance, rather than only on the plain value of the noise variance.

Fig. 5 shows the ratio between the condition numbers of the filtering matrices of the MMSE and ZF detectors for $1/\sigma_n^2 = 10$dB and $\text{cond}(\mathbf{H}) = 15$. The "Approximation" is obtained via our derived formula in (10). We remark that our approximation is accurate except for small values of $\sigma_{\min}$, where the error is tolerable. We also notice that the conditionality of $\mathbf{H}$ is highly improved at low $\sigma_{\min}$, where the filtering matrix tends to be orthogonal ($\text{cond}(\mathbf{W}_{\text{mmse}}) \approx 1.5$) at $\sigma_{\min} = 0.1$. The improvement by the MMSE regularization becomes negligible as $\sigma_{\min}$ increases.

Fig. 6 shows the ratio between the condition numbers of the filtering matrices of the MMSE and ZF detectors for $1/\sigma_n^2 = 30$dB and $\text{cond}(\mathbf{H}) = 15$. At this high SNR value, the improvement by the regularization imposed by the MMSE detector is also possible when $\sigma_{\min}$ is small. This explains the improvement in the BER for $(\sigma_{\min} > 0)$, as shown in Fig. 4.

## V. Conclusions

In this paper, we analyzed the performance of the linear ZF and MMSE detectors. Based on Weyl's Inequalities, we gave an accurate approximation of the ratio between the filtering matrices of the MMSE and ZF detectors. We revealed that the improvement in the conditionality of the channel matrix by the MMSE detector is dependent on the relationship between the noise variance and minimum singular values. Also, explicit formulas for the post-processing SNR for the ZF and MMSE detectors are derived, where we showed that the gain in the SNR by the MMSE detector increases with the increase in the number of transmit antennas. This is analytically argued by referring to the vanishing nature of the minimum singular value of the channel matrix as $N$ increases. Finally, simulations were carried out, where obtained results coincided with our analytical conjectures and approximate formulas.

## References

[1] R. Imer et al., "Multisite field trial for LTE and advanced concepts," *IEEE Communications Magazine,* vol. 47, no.2, pp. 92-98, Feb. 2009.
[2] E. Telatar, "Capacity of multi-antenna Gaussian channels," *European Transactions on Telecommunications,* vol. 10, pp. 585-595, Dec. 1999.
[3] W. Van Etten, "Maximum likelihood receiver for multiple channel transmission systems," *IEEE Transactions on Communications,* pp. 276-283, Feb. 1976.
[4] L. Barbero and J. Thompson, "Performance analysis of a fixed-complexity sphere decoder in high-dimensional MIMO systems," in *Proc. IEEE International Conference on Acoustics, Speech, and Signal Processing,* May 2006, pp. 557-560.
[5] K. J. Kim, J. Yue, R. A. Iltis, and J. D. Gibson, "AQRD-M/Kalman filter-based detection and channel estimation algorithm for MIMO-OFDM systems," *IEEE Transactions on Wireless Communications,* vol. 4, no. 2, pp. 710-721, Mar. 2005.
[6] P. Wolniansky, G. Foschini, G. Golden, and R. Valenzuela, "V-BLAST: An architecture for realizing very high data rates over the rich-scattering wireless channel," in *Proc. URSI International Symposium on Signals, Systems, and Electronics,* Oct. 1998, pp. 295-300.
[7] E. Agrell, T. Eriksson, A. Vardy, and K. Zeger, "Closest point search in lattices," *IEEE Transactions on Information Theory,* vol. 48, no. 8, pp. 2201-2214, Nov. 2002.
[8] D. Wubben, R. Bohnke, V. Kuhn, and K.-D. Kammeyer, "MMSE extension of V-BLAST based on sorted QR decomposition," in *Proc. IEEE Vehicular Telecomm. Conf.,* Oct. 2003, pp. 508-512.
[9] J. Wang and B. Daneshard, "A comparative study of MIMO detection algorithms for wideband spatial multiplexing systems," in *Proc. IEEE Wireless Comm. and Networking Conf.,* March 2005, pp. 408-413.
[10] J. Franklin, *Matrix Theory.* New Jersey: Courier Dover Publications, 2000.
[11] B. Hochwald, C. Peel, and L. Swindlehurst, "A vector-perturbation technique for near-capacity multiantenna multiuser communication - Part II: Perturbation," *IEEE Transactions on Communications,* vol. 53, no. 3, pp. 537-544, Mar. 2005.
[12] C. Meyer, *Matrix Analysis and Applied Linear Algebra.* Philadelphia: SIAM, 2001.
[13] A. Tulino and S. Verdu, *Random matrix theory and wireless communications.* The Netherlands, Delft: NOW, 2004.
[14] A. Edelman, "Eigenvalues and condition numbers of random matrices," Ph.D. Dissertation, MIT, Cambridge, 1989.